\documentclass{PoS}
\usepackage{lineno}
\usepackage{mathtools}
\usepackage{amsmath,amstext}

\title{Cosmic ray composition study using machine learning at the IceCube Neutrino Observatory}

\ShortTitle{Cosmic ray composition study using machine learning}

\author{The IceCube Collaboration\footnote{For collaboration list, see PoS(ICRC2019) 1177.}\\
{\itshape \href{http://icecube.wisc.edu/collaboration/authors/icrc19_icecube}{http://icecube.wisc.edu/collaboration/authors/icrc19\_icecube}}\\
E-mail: \email{matthias.plum@marquette.edu}
        }

\abstract{
The evaluation of mass composition of cosmic rays in the knee region ($\sim 3$~PeV) is critical to understanding the transition in the origin of cosmic rays from galactic to extragalactic sources.
The IceCube Neutrino Observatory at the South Pole is a multi-component detector consisting of the surface IceTop array and the deep in-ice IceCube detector. 
By applying modern machine-learning techniques to cosmic-ray air showers reconstructed coincidentally in both detector components of IceCube observatory, the energy and the mass of primary cosmic rays in this transition region can be measured.
In this contribution, we will discuss the reconstruction performance and composition sensitivity of IceCube observables presently under development.

\vspace{4mm}
{\bfseries Corresponding authors:}
\speaker{Matthias Plum}$^{1}$\\
{$^{1}$ \itshape Department of Physics, Marquette University, Milwaukee, WI, 53201, USA}
}
\FullConference{36th International Cosmic Ray Conference -ICRC2019-\\
		July 24th - August 1st, 2019\\
		Madison, WI, U.S.A.}

\newcommand{\lna}{$\langle\ln A\rangle$}
\newcommand{\s }{$S_{125}$}

\newcommand{\lte}{$\log_{10}(E/\mathrm{GeV})$}
\newcommand{\logdedx}{$\log_{10}(dE/dX_{1500\mathrm{m}})$}
\begin{document}

\section{Importance of Cosmic Ray Composition}\label{sec:intro}
High-energy cosmic rays consist mostly of charged particles.
Their composition is of key importance to understand their origin and acceleration processes. Additionally, the supposed transition from galactic to extragalactic sources of cosmic rays in the energy range from PeV to EeV should be visible in the evolution of the composition with energy. In air-shower physics, secondary particles are detected and it is presently only possible to measure the composition on a statistical basis.

\section{IceCube and IceTop Detectors}\label{sec:detector}
\begin{figure}[ht]
\centering
\begin{minipage}{0.49\textwidth}
\includegraphics[width=0.95\textwidth]{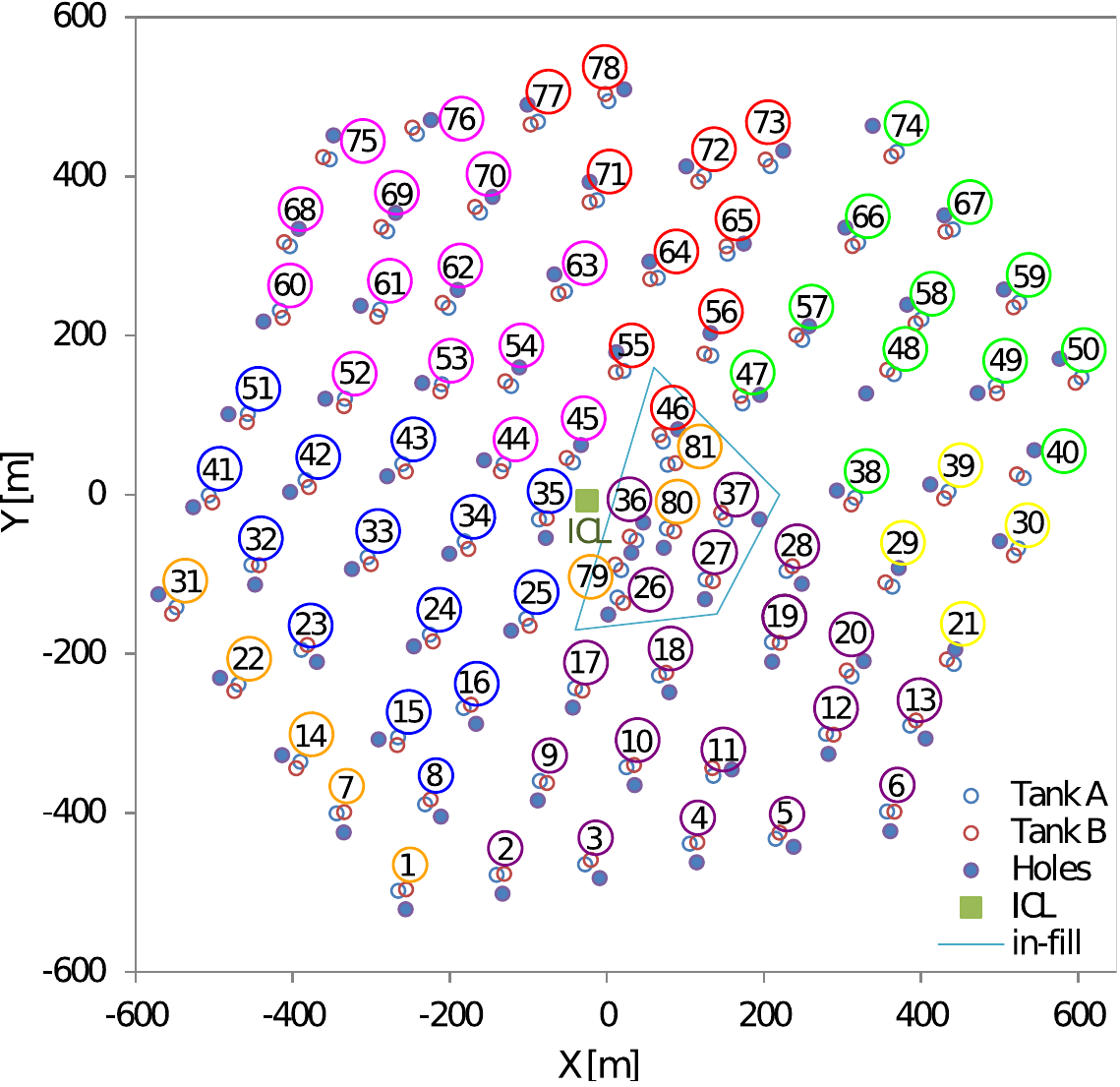}
\caption{\label{fig:IceTop} A top view of the IceTop surface array \cite{icetop_technical}. The color code correspond to deployment period of the IceCube string and the IceTop tanks.}
\end{minipage}\hfill
\begin{minipage}{0.49\textwidth}
\centering
\includegraphics[width=0.5\textwidth]{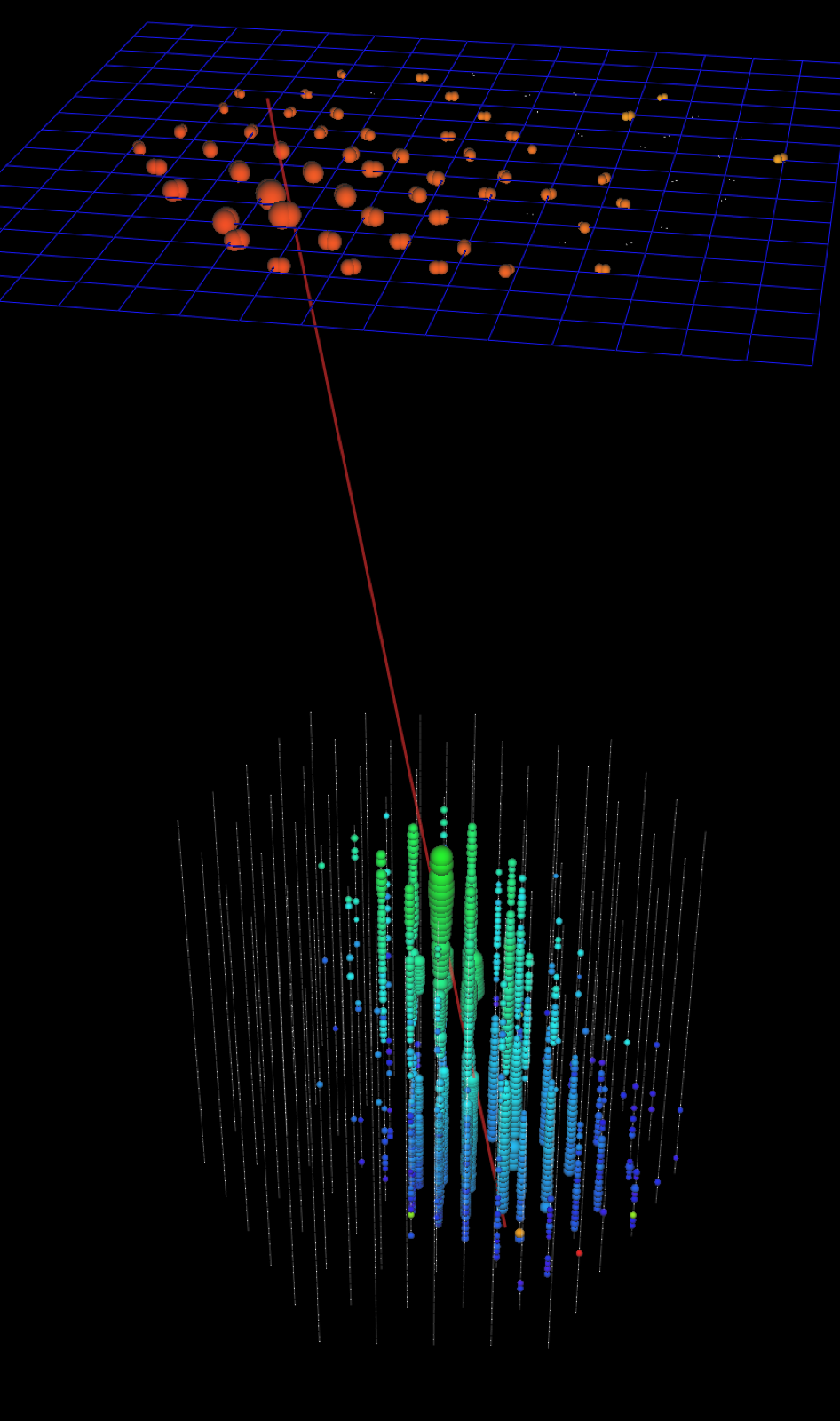}
\caption{\label{fig:Coincidence}Example coincidence event in IceCube and IceTop. The color represents the timing information from early (red) to late (blue). The size of the circles corresponds to the size of the measured signal in the PMTs. The red line is the reconstructed shower axis.}
\end{minipage}
\end{figure}

The IceCube Neutrino Observatory \cite{icecube_observatory} is a multi-purpose astroparticle detector located at the geographic South Pole. The detector is composed of the deep in-ice IceCube detector and the surface IceTop array. 
IceCube consists of 86 detector strings with 60 digital optical modules distributed between 1450~m and 2450~m beneath the surface of the ice.
The detector strings are arranged in a triangular grid with $\sim$125~m separation, as shown in Figure \ref{fig:IceTop}.

In total, IceCube covers an effective detector volume of 1~km$^3$. 
On top of most IceCube strings, two IceTop \cite{icetop_technical} tanks are placed, separated by roughly 10~m. 
Due to the construction of the detector over several years and the snow drift at the South Pole, the tanks are unevenly covered with snow, which is taken into account in the reconstruction and simulation of cosmic-ray events.
The deposited charge and the timing information of the cosmic-ray events are used to reconstruct the air-shower geometry and the lateral distribution of the charges at the surface and deep in-ice.  
This study includes only simulated cosmic-ray events which were reconstructed by the combination of both detector components in coincidence. An example coincidence event is shown in Figure \ref{fig:Coincidence}. Due to the geometric constraints of the IceCube and IceTop coincidence, only vertical ($\theta \leq 30^\circ$) events are used.
The simulations use the CORSIKA \cite{CORSIKA_Heck} air-shower generator, with FLUKA \cite{FLUKA:2006} as the low-energy hadronic interaction model and SIBYLL-2.1 \cite{Ahn:2009} as the high-energy interaction model. 
\section{Machine-Learning Reconstruction of the Primary Mass}\label{sec:mva}
Based on the successful application of machine-learning methods on 3 years of IceCube/IceTop coincidence data to reconstruct the primary energy and to derive a mass composition in \cite{Andeen:2019icrcw,IceCube_3year_composition_2019}, a further improvement of the mass composition resolution is studied by testing several new reconstructed detector observables for composition sensitivity.

\begin{figure}[ht]
\centering
\begin{minipage}{0.49\textwidth}
\includegraphics[width=1\textwidth]{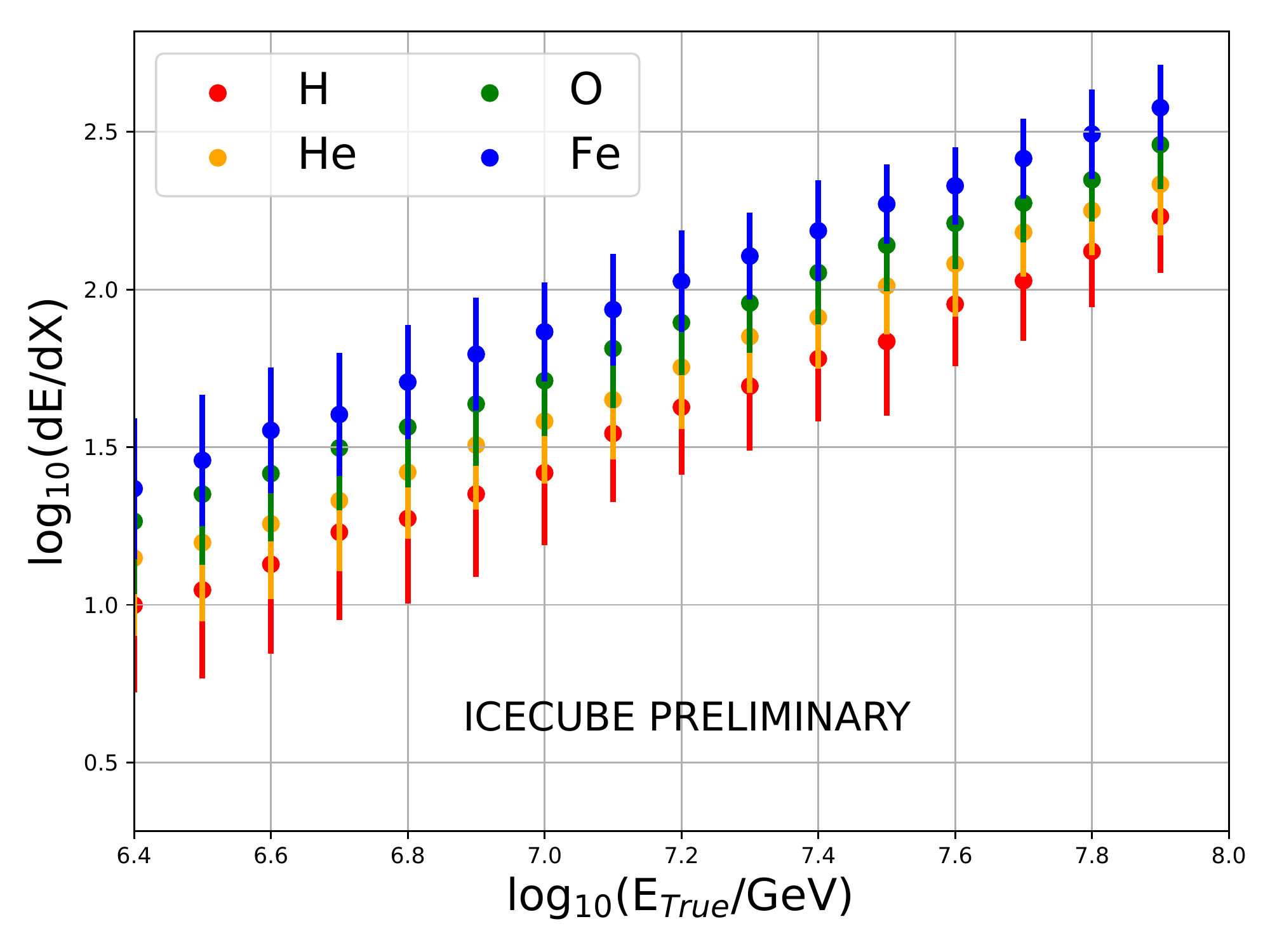}
\caption{\label{fig:dEdx}Mean and standard deviation of reconstructed \logdedx\,in IceCube as a function of energy for different primaries.}
\end{minipage}\hfill
\begin{minipage}{0.49\textwidth}
\includegraphics[width=1\textwidth]{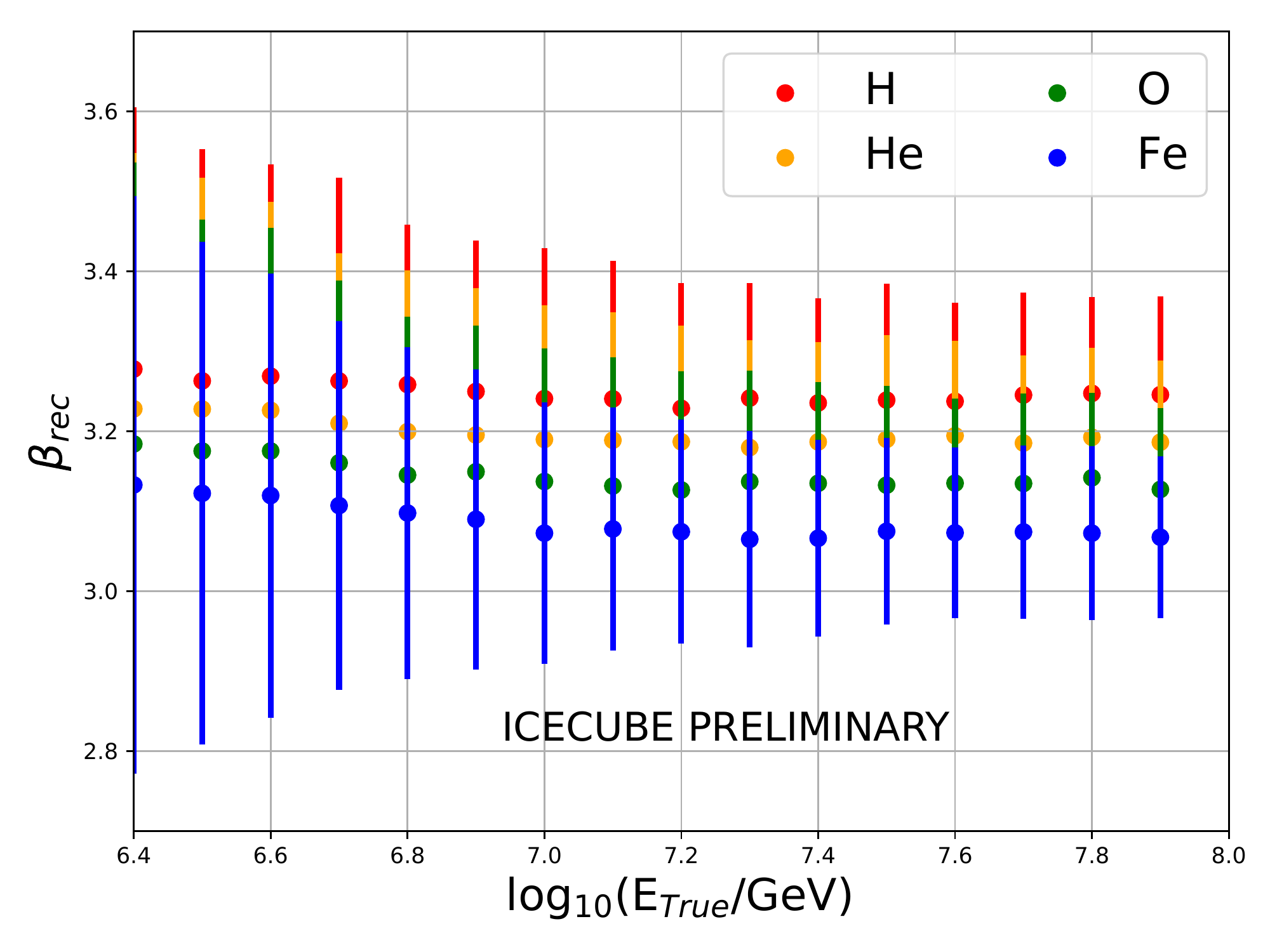}
\caption{\label{fig:beta} Mean and standard deviation of reconstructed $\beta$ in IceTop as a function of energy for different primaries.}
\end{minipage}
\end{figure}

The \textit{"baseline analysis"} of the simulation (similar to \cite{Andeen:2019icrcw,IceCube_3year_composition_2019}) is performed with the following observables: The logarithmic signal strength in IceTop at a distance of 125~m from the shower axis, \s, and the cosine of the reconstructed zenith angle cos$(\theta)$ derived from the standard reconstruction; from IceCube, the energy deposit of the high energy muon bundles at a slant depth of 1500~m \logdedx. \logdedx\, shows a strong primary composition dependence, which is shown in Figure \ref{fig:dEdx} for different primaries.
An \textit{"improved analysis"} is additionally performed to evaluate another composition sensitive variable, the shower age parameter $\beta$ from the `Double Logarithmic Parabola' fit \cite{icetop_technical} of the IceTop array, which also exhibits composition dependence over the whole energy range, shown in Figure \ref{fig:beta}.

The primary cosmic-ray elements used are proton, helium, oxygen and iron. Those elements are equally spaced over the \lna\,mass range and are therefore useful to train machine-learning regression algorithms. Due to the time-consuming full Monte Carlo production and detector reconstruction, roughly 2000 high quality full Monte Carlo simulation events per energy bin are used for the training and testing of a random forest tree (RFT) \cite{scikit-learn} using a 3-fold cross validation technique. Another independent $\sim2000$ full Monte Carlo simulation events per energy bin of \lte\,are used for verification of the machine-learning output, and for the final mass composition analysis.

\begin{figure}[ht]
\centering
\includegraphics[width=1.\textwidth]{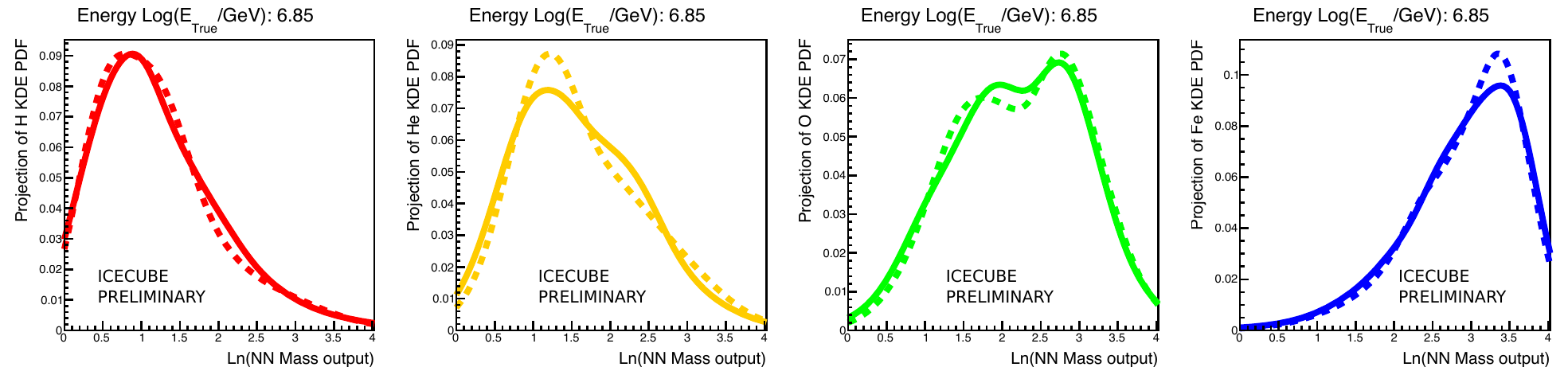}
\caption{\label{fig:template} 
Example bin with the KDE \cite{KDE_Cramer} mass template PDFs generated with a RFT regressor of the baseline (solid) and improved (dashed) analysis in the energy bin \lte=6.8-6.9.}
\end{figure}

The machine-learning output of the verification sample is converted into kernel density estimation (KDE) \cite{KDE_Cramer} probability density functions (PDF), which are used as templates for every energy bin and individual elementary groups using the RooFit toolkit \cite{roofit_2003} as in \cite{Andeen:2019icrcw,IceCube_3year_composition_2019}. An example energy bin is shown in Figure \ref{fig:template}.
The template shape of the new improved (dashed) templates are distinctively different from the templates of the baseline (solid) analysis.
These template PDF's for each energy bin of the four elementary groups are combined to a joined mass composition response PDF by introducing a weight factor for each primary given by

\begin{equation*}
    P_{mass}(X) = \sum_{i=H,He,O,Fe} w_i \cdot P_i(X) \text{ with }
    \sum_{i=H,He,O,Fe} w_i = 1,
\end{equation*}
where $X$ is the mass output of the machine learning. Due to the constrained weight factors $w_i$, the results for the elementary groups are highly correlated with each other.
On a statistical basis, $P_{mass}$ is used to fit the data to determine the contribution of each elementary primary group with an extended likelihood fit. 

\section{Reconstruction Capabilities of Different Composition Scenarios}\label{sec:fraction_reconstruction}
The reconstruction and the improvement of the mass resolution of the template method is studied for various scenarios and both cases, with and without $\beta$, are compared to each other.
The mass composition response PDF for each energy bin can be used to generate fast mock scenario data sets where the number of events and the fractions of each primary group are artificially selected.
In this way the reconstruction capabilities of the method are tested for several Monte Carlo scenarios.

\begin{figure}[ht]
\centering
\includegraphics[width=0.33\textwidth]{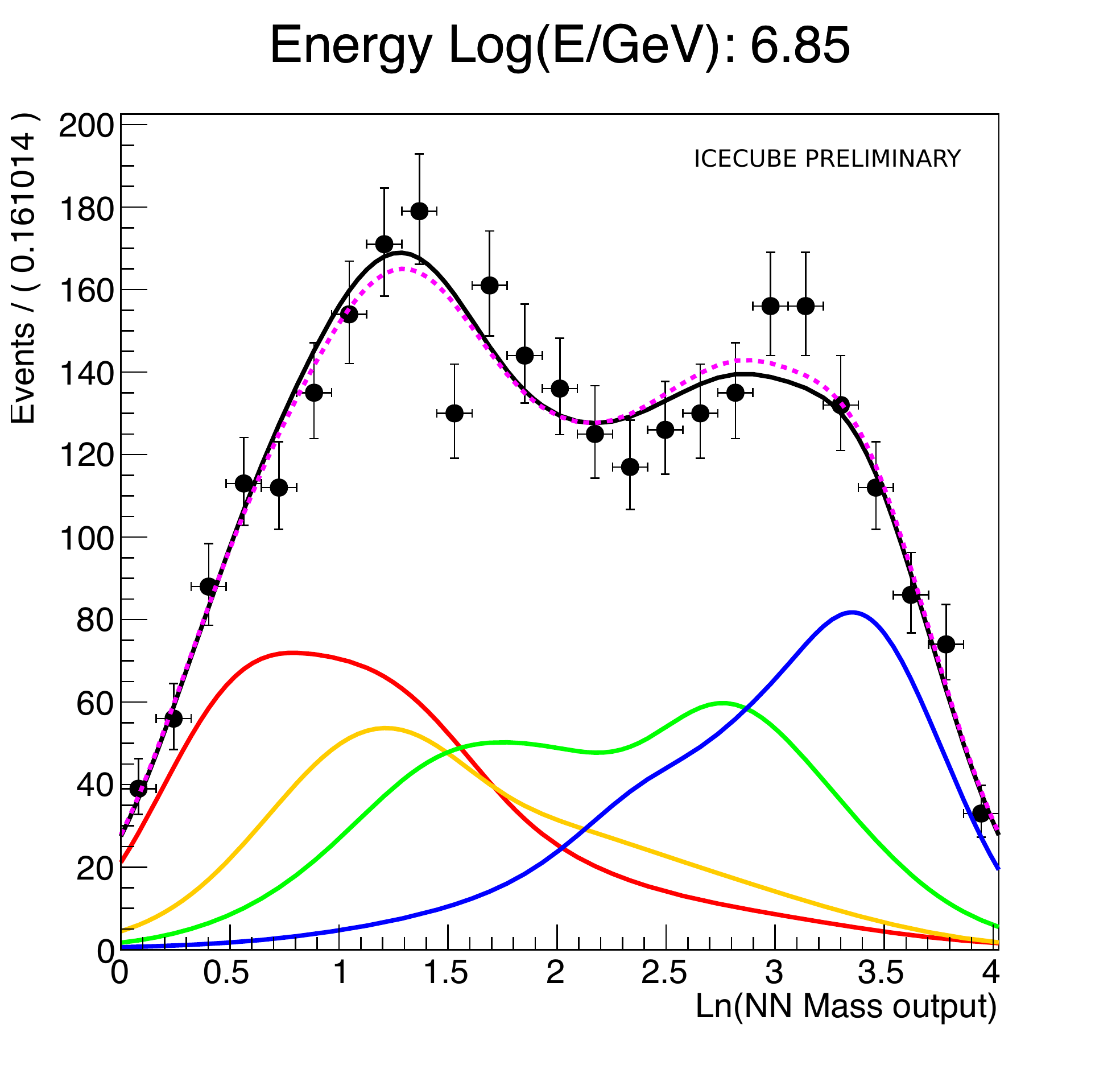}
\caption{\label{fig:testframe} Example fast Monte Carlo data set from the improved analysis from the maximum mixing scenario of the four elementary groups (Fraction: 25\%:25\%:25\%:25\%) in the energy bin \lte=6.8-6.9 with in total 3000 MC events. The generator PDF is shown as a solid black line, the fitted PDF is shown as a dashed magenta line. The corresponding weighted elementary group PDF are shown in red for H, yellow for He, green for O and blue for Fe.}
\end{figure}

\begin{figure}[ht]
\centering
\includegraphics[width=1\textwidth]{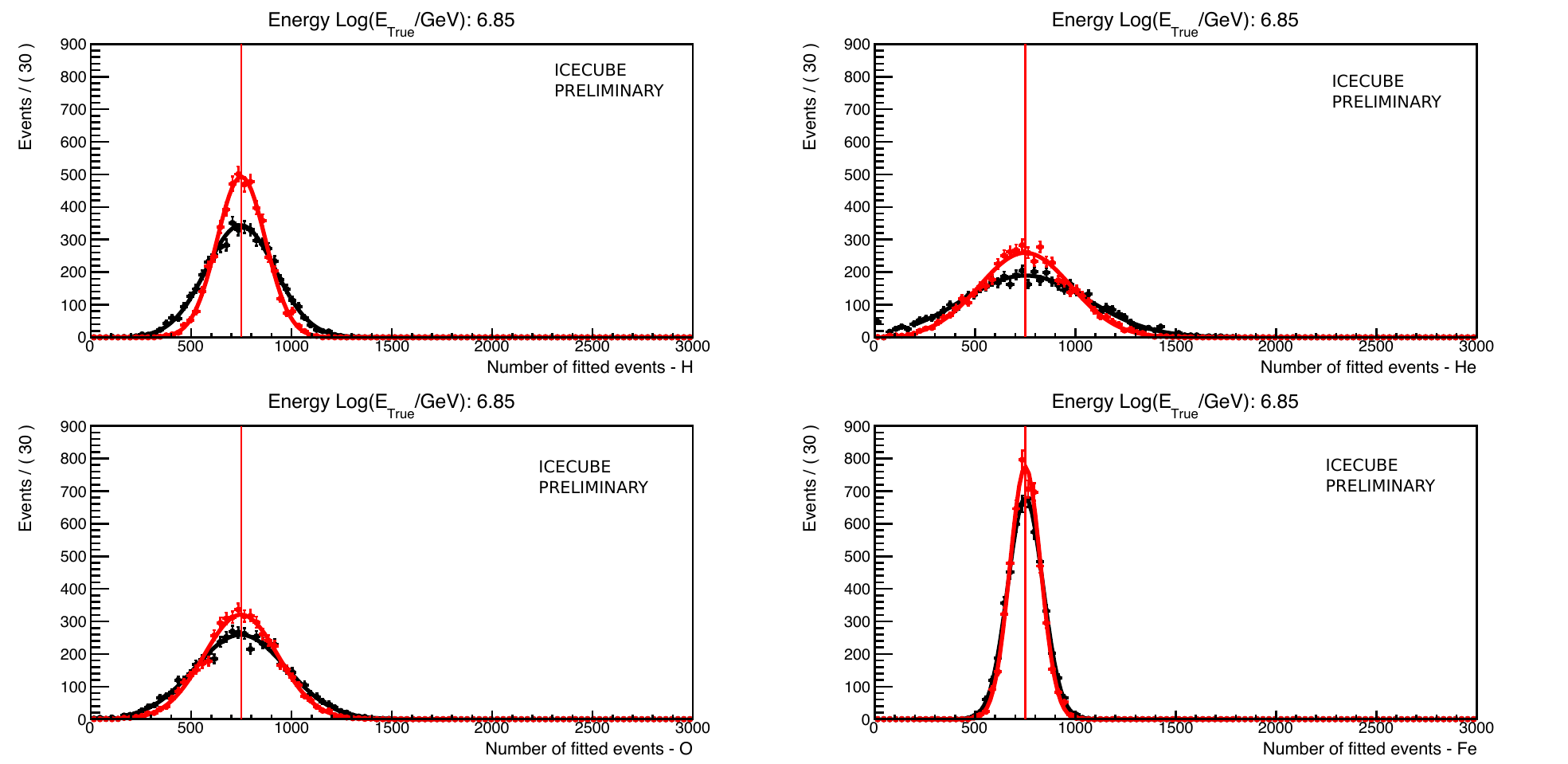}
\caption{\label{fig:MC_study}Example bin of the reconstructed number of events from the Monte Carlo reconstruction study for the maximum mixing scenario of the four elementary groups (Fractions: 25\%:25\%:25\%:25\%) in the energy bin \lte=6.8-6.9 with in total 3000 MC events per bin. The baseline analysis results are shown in black, the improved ones are shown in red. A Gaussian fit was applied to each distribution to measure the average reconstructed fraction and the resolution.
The vertical line shows the Monte Carlo truth in this scenario.}
\end{figure}

\begin{figure}[ht]
\centering
\includegraphics[width=1\textwidth]{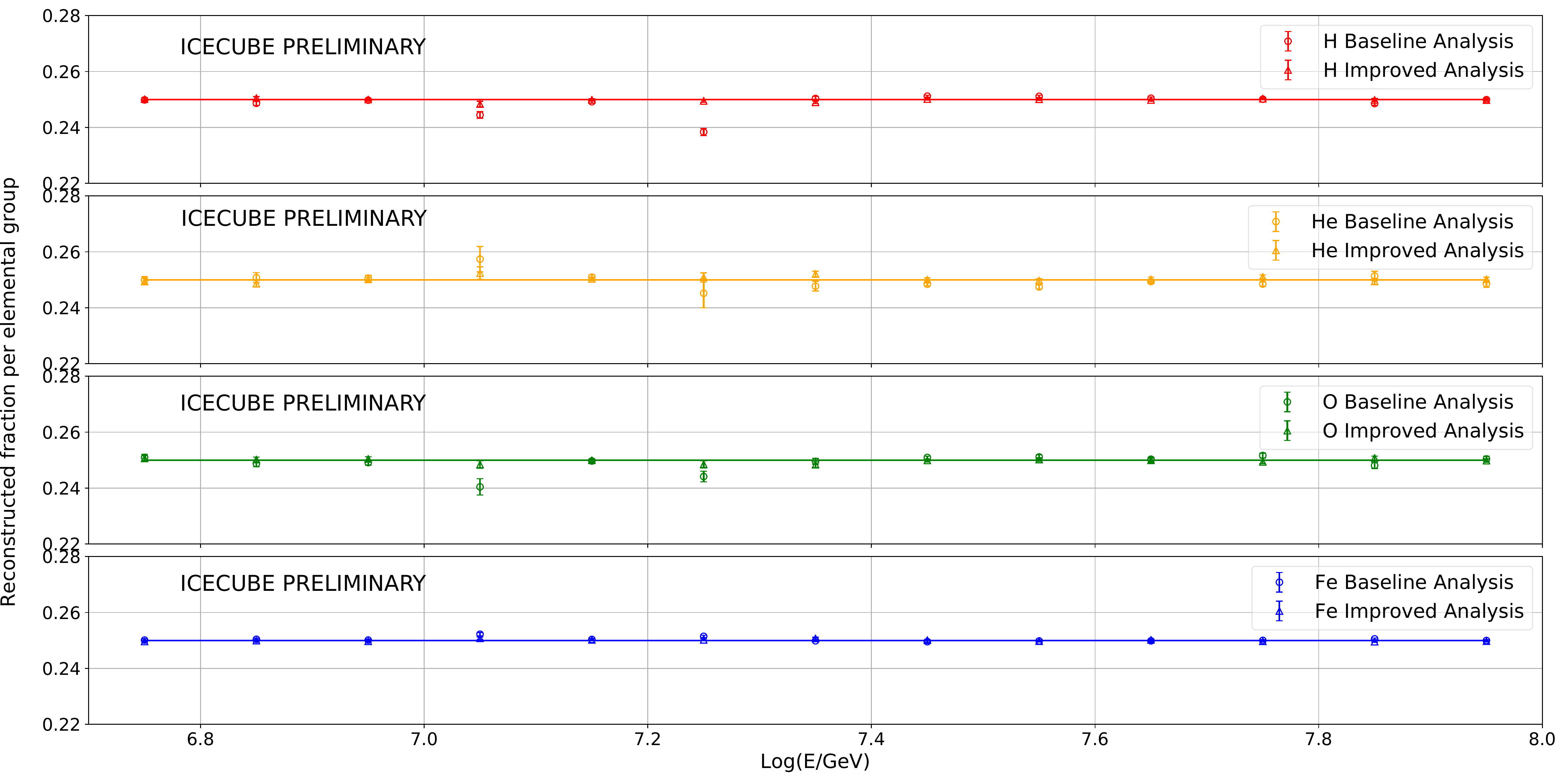}
\caption{\label{fig:MVA_fraction_comparison} Comparison of average reconstructed mass fractions by the baseline and improved analysis with each bin containing 3000 events. Both analyses reconstruct on average the Monte Carlo truth, represented by the solid line. Note the zoom on the y-axis around 25\% to emphasize the very minor change in results.}
\end{figure}

\begin{figure}[ht]
\centering
\includegraphics[width=1\textwidth]{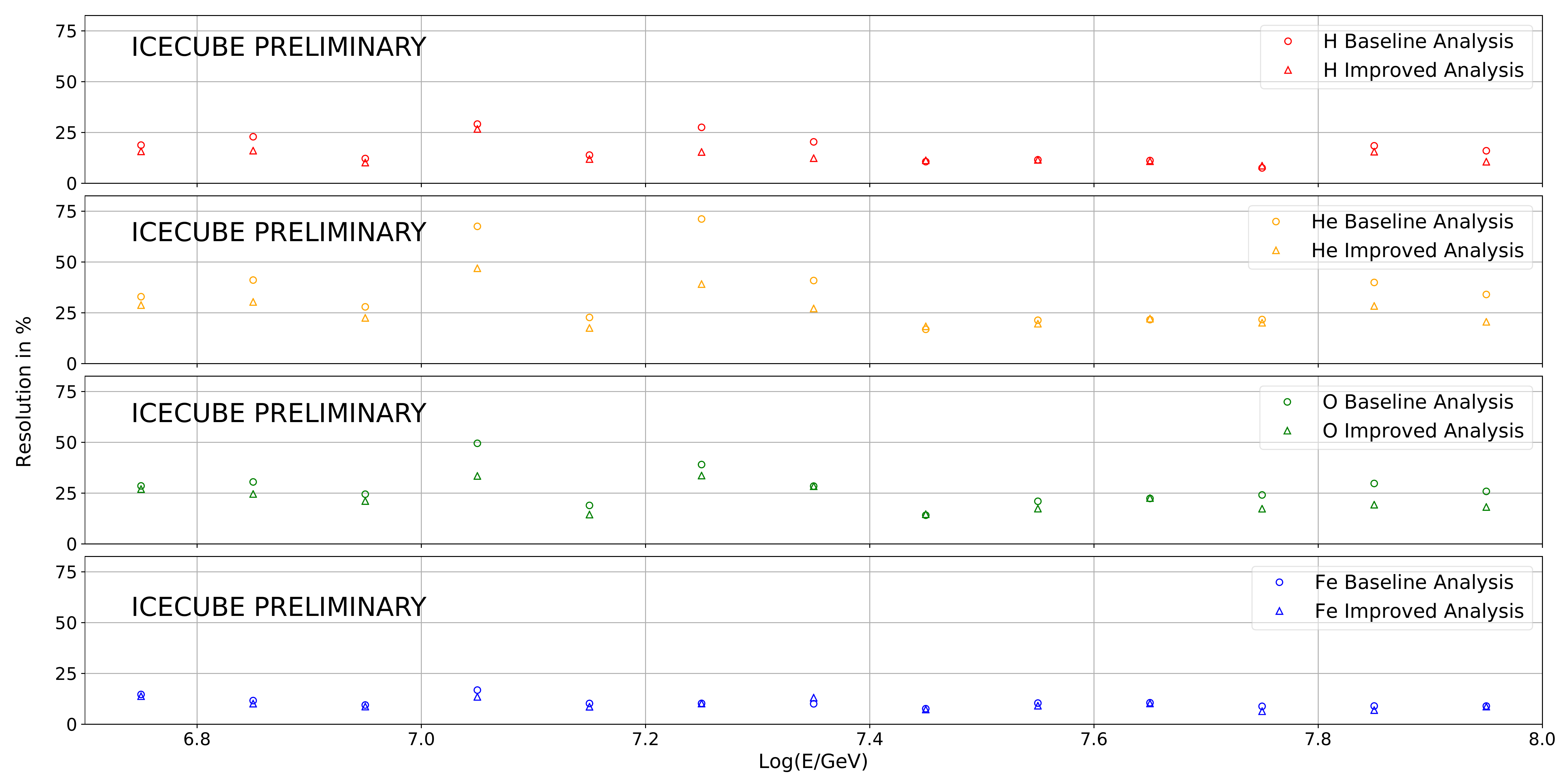}
\caption{\label{fig:MVA_resolution_comparison} Comparison of mass composition resolution derived from a Monte Carlo study with 3000 MC events per bin. The improved analysis shows a slight improvement over the baseline analysis.}
\end{figure}

The first tested scenario is the maximum mixing scenario of the four elementary groups (Fractions: 25\%:25\%:25\%:25\%). A generated Monte Carlo data set with 3000 total events and the corresponding fit is shown in Figure \ref{fig:testframe}. The input PDF is reconstructed within the statistical uncertainties.
This procedure is repeated several thousand times and the fitted number of events per elementary group are collected and analyzed. An example energy bin of this Monte Carlo study is shown in Figure \ref{fig:MC_study}, where the distribution of the fit parameter of the baseline is slightly broader than for the improved analysis. 
The average reconstructed fractions and the mass resolutions for each elementary group are measured using a Gaussian fit to these distributions.
This study is repeated for all other energy bins with the same parameters. 
The mass fraction in this scenario is on average accurately reconstructed as shown in Figure \ref{fig:MVA_fraction_comparison} for the whole energy range.
Both the baseline and the improved analysis reconstructed the true composition with high accuracy inside the statistical uncertainties.
The comparison of the mass resolution from the baseline and the improved analysis is shown in Figure \ref{fig:MVA_resolution_comparison},
which shows that the additional information of a new composition sensitive variable improves the mass resolution over the whole energy range.
The fractions of the intermediate element groups of helium and oxygen are intrinsically uncertain due to the large overlap with their neighboring distributions and are showing a slightly larger improvement inside the statistical uncertainties of the fit method than the proton an iron groups.

\begin{figure}[ht]
\centering
\includegraphics[width=1\textwidth]{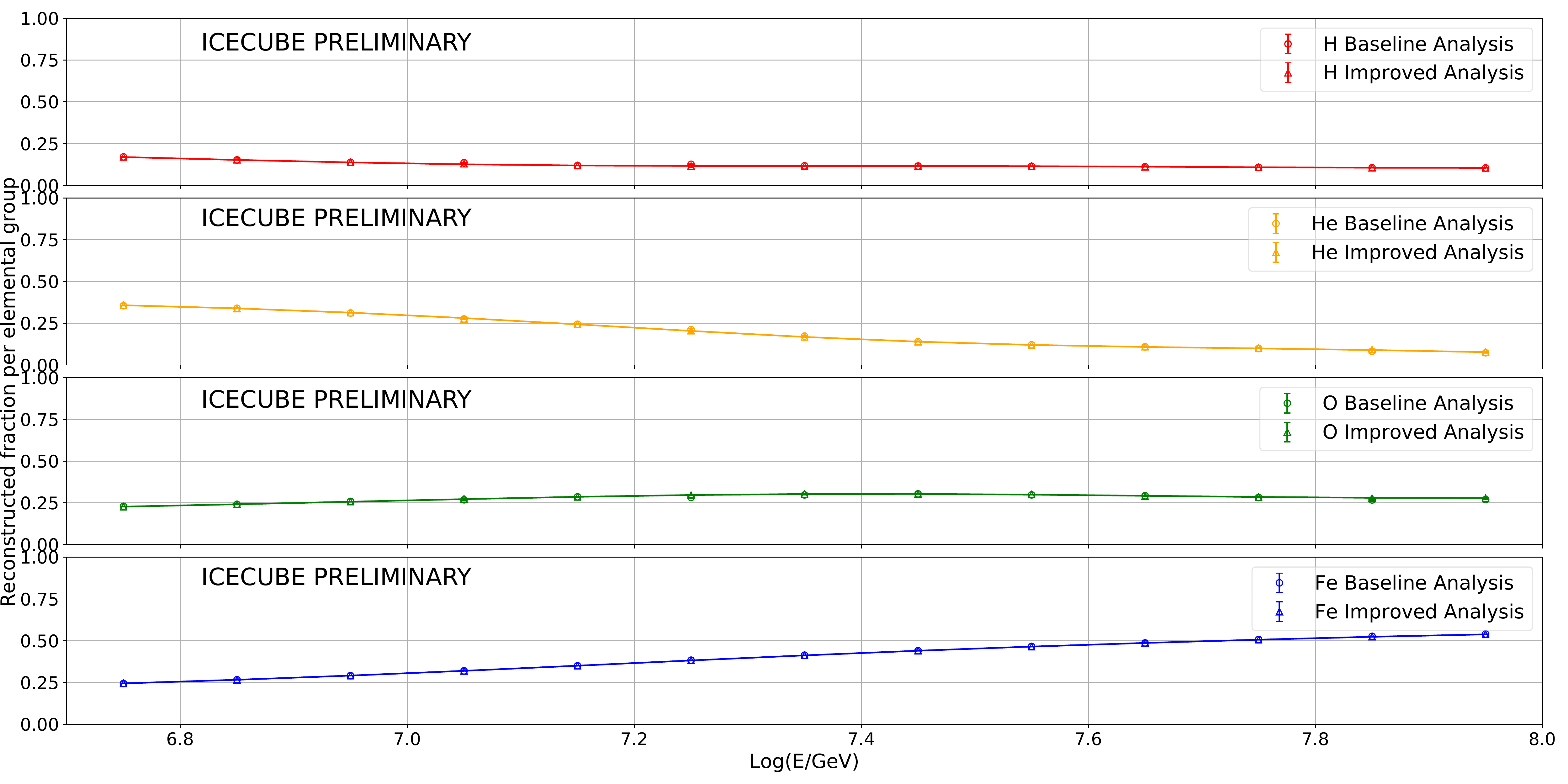}
\caption{\label{fig:H4a_comparison} Comparison of mass reconstruction based on the H4a \cite{Gaisser_H4a} model of the baseline and the improved analysis with 3000 MC events per bin.}
\end{figure}

A realistic cosmic-ray scenario assuming an H4a \cite{Gaisser_H4a} composition is also tested in this energy range. The reconstructed average composition is shown in Figure \ref{fig:H4a_comparison}.   
The plot shows that both analyses are capable of reconstructing the primary mass composition in a realistic source scenario.

\section{Summary \& Outlook}\label{sec:summary}
The mass composition template analysis of the IceCube/IceTop coincident events is verified with Monte Carlo scenarios. Full Monte Carlo simulation was used to train and test a random forest tree regressor and elementary group probability density templates for every energy bin were created. The templates were used to generate fast Monte Carlo data sets for both a maximum mixing and an H4a\cite{Gaisser_H4a} input scenario.
The mass resolution is measured by generating and fitting several thousand fast Monte Carlo data sets and analyzing the fit results. A comparison of the mass resolution between the baseline analysis used in \cite{Andeen:2019icrcw,IceCube_3year_composition_2019} and the future improved analysis using the shower age parameter $\beta$ is presented and shows a slight improvement over the whole energy range. 

In the future, an improved coincidence reconstruction \cite{Dvorak:2019icrcw} will add several new composition sensitivity variables like the air shower curvature and the muon density at the surface.
The joint measurements of the proposed scintillator array \cite{kauer:2019icrc} and IceTop will add information sensitive to composition by the inclusion of the shower parameter as shown in \cite{agnes:2019icrc}.
Also, the proposed IceAct \cite{Auffenberg:2019icrcw} array will provide additional air shower information about the electromagnetic shower component like the center-of-gravity, which will further improve cosmic-ray measurements of the composition and also produces opportunities to investigate and constrain hadronic interaction models.

\clearpage
\bibliographystyle{ICRC}
\bibliography{references}

\end{document}